\documentclass[twocolumn,showpacs,amsmath,amssymb]{revtex4}

\usepackage{graphicx}
\usepackage{dcolumn}
\usepackage{bm}
\begin{document}

\title{Supercurrent in Long SFFS Junctions with Antiparallel Domain
Configuration}

\author{Ya.~M.~Blanter$^{a}$ and F.~W.~J.~Hekking$^{b}$}
\affiliation{ $^{a}$Department of NanoScience and DIMES, Delft
University of
Technology, Lorentzweg 1, 2628 CJ Delft, The Netherlands\\
$^{b}$Laboratoire de Physique et Mod\'elisation des Milieux
Condens\'es, CNRS \& 
Universit\'e Joseph Fourier, B.P. 166, 38042 Grenoble cedex 9, France}
\date{\today}

\begin{abstract}
We calculate the current-phase relation of a long Josephson junction
consisting of two ferromagnetic domains with equal, but opposite
magnetization $h$, sandwiched between two superconductors. In the clean
limit, the current-phase relation is obtained with the help of
Eilenberger equation. In general, the supercurrent oscillations are  
non-sinusoidal and their amplitude decays algebraically when the
exchange field is increased. If the two domains have the same size,
the amplitude is independent of $h$, due to an exact cancellation of
the phases acquired in each ferromagnetic domain. These results change
drastically in the presence of disorder. We explicitly study two 
cases: Fluctuations of the domain size (in the framework of the
Eilenberger equation) and impurity scattering (using the Usadel
equation). In both cases, the current-phase relation becomes
sinusoidal and the amplitude of the supercurrent oscillations is 
exponentially suppressed with $h$, even if the domains are identical
on average. 
\end{abstract}

 \maketitle

\section{Introduction}

Hybrid systems containing superconducting and ferromagnetic
elements gained recently a lot of attention due to experimental
progress as well as possible applications in magnetoelectronics
and quantum information. Theoretical studies are revealing a
variety of new features, making these system generators of novel
theoretical concepts.

It is a common knowledge that current in hybrid normal metal --
superconductor (NS) systems flows by means of Andreev reflections:
an electron in N is reflected from the NS interface as a hole with
the opposite charge and velocity. Imagine first that the piece of
normal metal is ballistic. An electron at the Fermi surface is
reflected as a hole at the Fermi surface, and they propagate in
the normal metal with the same phase. If the electron is taken at
a finite energy $E$ (counted from the Fermi surface), a momentum
mismatch $\delta p = 2E/v_F$ between this electron and the
reflected hole appears, $v_F$ being the Fermi velocity.

Consider now an interface between an ($s$-wave) superconductor and a
ferromagnet. 
Electron and hole have opposite spin directions, and the exchange
field $h$ in the ferromagnet leads to a Zeeman splitting of energies
of the two different spin projections. Thus, even an electron and a
hole at the Fermi surface acquire the momentum mismatch $2h/v_F$;
hence their relative phase grows as $\delta \varphi = 2hx/(\hbar
v_F)$, where $x$ is the distance from the interface. This affects 
phase-sensitive physical quantities like the supercurrent in
superconductor -- ferromagnet -- superconductor (SFS) junctions: It
becomes an oscillating function of the thickness $d$ of the
ferromagnetic layer, with a period $\hbar v_F/2h$. If, furthermore,
the ferromagnet is diffusive, the oscillating behavior is 
accompanied by an exponential decay $\propto \exp ( -(h/\hbar D)^{1/2}
d)$, where $D$ is the diffusion coefficient. Typically, $h$ is much
larger than the superconducting gap $\Delta$, and thus the length
scales related to the magnetic field are much shorter 
than the superconducting coherence length $\xi$, $\hbar v_F/\Delta$
and $(\hbar D/\Delta)^{1/2}$ in the clean and diffusive case,
respectively. In other words, the proximity effect is suppressed in
the ferromagnet.

This qualitative discussion suggests that the main effect observed
in SFS contacts is oscillations of the supercurrent with the
thickness of the ferromagnetic layer --- the transition from a
so-called $0$-state (energy of the contact is minimum for zero
phase difference between the superconductors) to a $\pi$-state
(energy is minimum for a phase difference $\pi$). This topic was
at the focus of attention since the early exploration of the
field~\cite{Bulaevskii77}. Theoretically, the $\pi$-state was
predicted in a variety of SFS junctions:
Ballistic~\cite{Bulaevskii,Radovic00,Chtchelkatchev,Bergeret2},
short diffusive~\cite{Buzdin,Demler}, long
diffusive~\cite{Buzdin,Heikkila,Bergeret2}, ferromagnetic
insulating barrier~\cite{Bulaevskii77,Tanaka,Fogelstrom},
ballistic~\cite{Bergeret1,Barash1,Barash2} and
diffusive~\cite{Koshina1,Bergeret2,Koshina2,Krivoruchko,Golubov1,Golubov2}
junctions with a barrier separating two ferromagnetic layers, and
ballistic~\cite{Radovic03} and diffusive~\cite{Golubov2} with two
tunnel barriers. The transition to the $\pi$-state was recently
observed experimentally in SFS
junctions~\cite{Ryazanov,Kontos,Guichard,Blum}. All these
observations are limited to small thickness of ferromagnetic
layer(s), $d \lesssim (\hbar D/h)^{1/2}$. For thicker layers,
supercurrent does not exist.

In this situation, it is useful to understand how one can enhance
the proximity effect. Several options have been recently discussed
in the literature. First, the above qualitative argument assumes
that the pairing between an electron and a hole participating in
the Andreev reflection is {\em singlet} -- they have opposite spin
projections. Obviously, if the superconductor allows for a
non-trivial symmetry of the order parameter, this needs not be the
case, and {\em triplet} pairing between an electron and a hole
with the same spin projection can arise. Since triplet-paired
electron and hole at the Fermi surface have no momentum
difference, they can propagate with the same phase and enhance the
proximity effect. Coupling of two $d$-wave superconductors via a
ferromagnetic layer has been considered in
Ref.~\onlinecite{Radovic99}. Moreover, triplet pairing can even
appear in a contact of an $s$-wave superconductor and a
ferromagnet, provided the magnetization in the latter is
non-uniform~\cite{Larkin,Bergeret3,Kadigrobov}. In this case, the
proximity effect survives at the same distance $\xi$ from the
interface as in non-magnetic metals. Indeed, the supercurrent in
SFS junctions with non-uniform magnetization is considerably
enhanced~\cite{Bergeret2}. We also mention that the supercurrent
in a long diffusive SFS junction is exponentially suppressed only
{\em on average}; phenomena related to the proximity effect still
occur in such a junction as a result of {\em  mesoscopic
fluctuations} around average quantities~\cite{Spivak}. Finally, if
the ferromagnetic layer is split into domains, the coherence can
be preserved if an electron and a hole propagate between the
superconducting electrodes along the two sides of a domain
wall~\cite{Melin}.

In this Article, we explore a different way to enhance the
supercurrent in SFS junctions. Imagine first that the junction is
ballistic and the ferromagnetic layer consists of two domains with
opposite directions of the magnetization, as shown in Fig.~1.
Triplet pairing is not generated in this geometry. Consider an
electron and an Andreev-reflected hole propagating from left to
right between the superconducting electrodes. They first acquire
the relative phase $\delta \varphi_1 = 2hx_1/(\hbar v_F)$, $x_1$
being the distance traversed in the first ferromagnetic layer.
However, in the second layer the exchange field has the opposite
sign, and the phase gain $\delta \varphi_2 = -2h x_2/(\hbar v_F)$
partially compensates $\delta \varphi_1$. For $x_1 = x_2$ we have
full compensation: The ferromagnetic bilayer behaves as a piece of
normal (not ferromagnetic) metal, and the proximity effect is
fully restored. Indeed, previous studies of SFS contacts where the
two ferromagnetic domains were separated by a barrier, found that
the supercurrent in the antiparallel domain configuration is
enhanced with respect to the parallel
one~\cite{Bergeret1,Koshina2,Barash2,Golubov2}. If the domains are
identical, there is no transition to the $\pi$-sate in the
antiparallel configuration.

Below, we consider such a situation quantitatively.
Section~\ref{cleancont} treats a ballistic SFFS junction with two
ferromagnetic domains parallel to the superconducting interfaces.
We show that this system behaves as a ballistic SFS junction with
an {\em effective} exchange field. If the widths of the two
domains are the same, this effective field vanishes. In the next
two Sections, we study the effect of disorder in the same system
and show that supercurrent in diffusive SFFS junctions decays
exponentially with their width, similarly to SFS contacts without
domains. We consider long junctions, $d \gg \xi$, and assume that
the superconducting electrodes do not influence the magnetic
structure of the contact.

\section{Clean SFFS contact}
\label{cleancont}

We consider first a system of two clean ferromagnetic
strips~\cite{foot1} with the thicknesses $d_1$ and $d_2$ and
antiparallel orientations located between two superconductors
(Fig.~1). The dynamics of quasiparticles in this system are
described by the Eilenberger equation,
\begin{equation} \label{Eilenb}
-i v_F \bm{n} \bm{\nabla} \breve{g}_{\sigma} (\bm{r}, \bm{n}) =
\left[
 (i\omega \mp h\sigma) \breve{\tau}_3 + \breve{\Delta},
 \breve{g}_{\sigma} (\bm{r}, \bm{n}) \right]_{-} .
\end{equation}
Here the semi-classical Green's function $\breve{g}_{\sigma}$ is a
matrix in Nambu space,
\begin{displaymath}
\breve{g}_{\sigma} = \left( \begin{array}{lr} g_{\sigma} &
f_{\sigma}
\\ f^+_{\sigma} & -g_{\sigma} \end{array} \right),
\end{displaymath}
which describes the singlet pairing (the triplet component is not
generated in our geometry), and the spin index $\sigma = \pm 1$.
The exchange field $h$ is zero in the superconducting banks, and
has antiparallel orientations in the ferromagnets: The upper/lower
signs in Eq.~(\ref{Eilenb}) corresponds to the left/right
ferromagnet ($h > 0$). To stay in the framework of the
semi-classical consideration, we have assumed that the Zeeman
splitting $h$ is much weaker than the Fermi energy, but can be
arbitrary in comparison with the superconducting gap $\Delta$. We
put $\hbar = 1$; it will be restored in the final results.

\vspace{1.4cm}

\begin{figure}[ht]
\includegraphics[angle=0,width=7.cm]{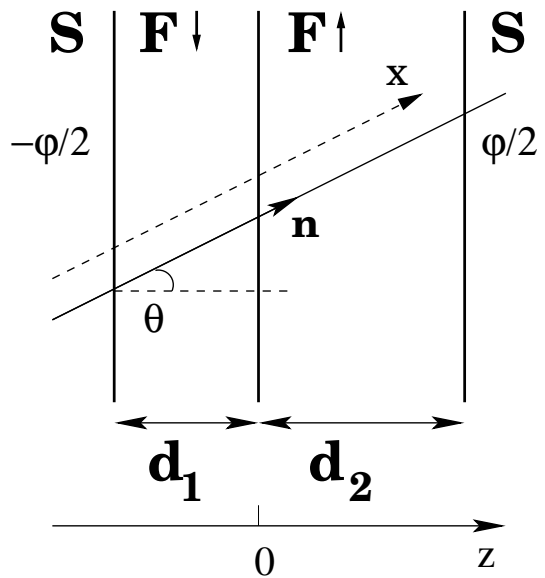}
\caption{\label{fig1} Setup with two ferromagnetic domains with
antiparallel configurations.}
\end{figure}

In this Article, we consider the case of a {\em long contact}: The
thicknesses of both ferromagnetic layers are much larger than the
superconducting coherence length, $d_{1,2} \gg \hbar v_F/\Delta$.
Then the matrix $\breve{\Delta}$ can be taken in a piecewise
approximation: It is zero in both ferromagnets, and
\begin{displaymath}
\breve{\Delta} = \left( \begin{array}{lr} 0 & \Delta e^{i\chi}
\\ - \Delta e^{-i\chi} & 0 \end{array} \right)
\end{displaymath}
in the superconductors. Here $\chi = -\varphi/2$ and $\chi =
\varphi/2$ in the left 
and the right superconducting bank, respectively.

In the bulk superconductor far from the contacts the Green's
function is isotropic and equals for $\vert \omega \vert < \Delta$
\begin{eqnarray} \label{boundcond1}
\breve{g}_{\sigma}^{bulk} = \frac{1}{\sqrt{\Delta^2 + \omega^2}}
\left(
\begin{array}{lr} \omega & -i\Delta e^{i\chi} \\ i\Delta e^{-i\chi} &
-\omega \end{array} \right).
\end{eqnarray}
In addition, the Green's function and its derivative must be
continuous at each interface.

We introduce the coordinate $x$ parallel to $\bm{n}$ and directed
from left to right. Let us choose $x=0$ at the boundary of the
left superconductor; then $x=d_1/\cos \theta$ at the interface of
the two ferromagnets, and $x = (d_1 + d_2)/\cos \theta$ at the
boundary of the right superconductor. The quasiparticles in the
clean system move along a straight line (Fig.~\ref{fig1}). It
follows from Eq.~(\ref{Eilenb}) that the normal component
$g_{\sigma} (\bm{r}, \bm{n})$ is constant along the trajectory
inside the ferromagnets. The calculation gives
\begin{equation} \label{clean1}
g_{\sigma} (\bm{n}) = \frac{\sqrt{\Delta^2 + \omega^2} \sin \alpha
+ i\omega (1 + \cos \alpha)}{\omega \sin \alpha + i \sqrt{\Delta^2
+ \omega^2} (1 + \cos \alpha)},
\end{equation}
where the phase $\alpha$ accumulated along the trajectory is
\begin{equation} \label{phase1}
\alpha = \frac{2i\omega}{v_F}\frac{d_1 + d_2}{\cos \theta} +
\frac{2 h\sigma}{v_F} \frac{d_1 - d_2}{\cos \theta} - n_x \varphi,
\ \ \ n_x = \pm 1.
\end{equation}

The supercurrent density is expressed as follows,
\begin{equation} \label{supercur1}
j = -i\pi e v_F \nu \sum_{\sigma} T \sum_{\omega} \int d\bm{n}
g_{\bm{n}} {\bm n},
\end{equation}
where $\nu$ is the density of states. For $h=0$ Eq.~(\ref{supercur1})
gives the supercurrent of a long clean SNS (non-ferromagnetic)
junction, as considered in Ref.~\onlinecite{Svidzinsky}, which we
follow in the general case. The expression is even in $\omega$; for
zero temperature (case of interest here) the summation can be replaced
by an integration over frequencies. We subsequently introduce a new 
integration variable $\omega = \Delta \sinh u$ and arrive at the
intermediate expression
\begin{eqnarray} \label{supercur2}
& & j = 2ev_F\nu \Delta \sum_{\sigma} \int_0^{\infty} du \cosh u
\int_0^{\pi/2} d\theta \cos \theta \\
& & \times \mbox{Im} \tanh \left[u + \Delta \sinh u \frac{d_1 +
d_2}{v_F \cos \theta} + ih\sigma \frac{d_1 - d_2}{v_F \cos \theta}
+ i \frac{\varphi}{2} \right]. \nonumber
\end{eqnarray}
For long contacts, $\Delta d_{1,2} \gg \hbar v_F$, the first term
in the argument of the hyperbolic tangent can be disregarded.
Using the identity
\begin{displaymath}
\mbox{Im} \tanh y = 2 \sum_{k=1}^{\infty} (-1)^k \ \mbox{Im} \
e^{-2ky},
\end{displaymath}
we obtain the final expression for the supercurrent,
\begin{eqnarray} \label{supercur3}
j & = & \frac{4ev_F^2 \nu\hbar}{d_1 + d_2} \sum_{k=1}^{\infty}
\frac{(-1)^{k+1}}{k} \sin k\varphi \nonumber \\
& \times & \int_1^{\infty} \frac{dx}{x^3 \sqrt{x^2-1}} \cos
\frac{2kh(d_1-d_2)x}{v_F \hbar}.
\end{eqnarray}

For $h=0$, we return to the clean long SNS contact,
\begin{equation} \label{supSNS1}
j = j_0 \sum_{k=1}^{\infty} \frac{(-1)^{k+1}}{k} \sin k\varphi,
\end{equation}
where $j_0 = \pi ev_F^2 \nu\hbar/(d_1 + d_2)$. This describes the
well-known sawtooth current-phase relation found earlier in
Ref.~\onlinecite{Ishii}.

\begin{figure}[ht]
\includegraphics[angle=0,width=7.cm]{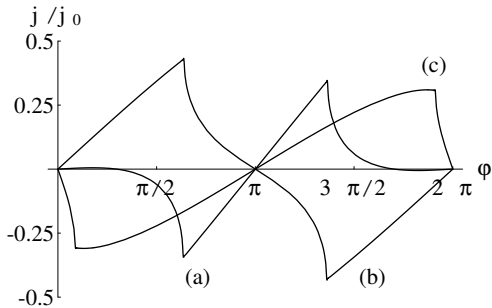}
\caption{\label{fig2} Supercurrent $j$ in units of $j_0$ as a
function of $\varphi$ for various values of $h$: $h\vert d_1 -
d_2\vert /\hbar v_F$ = 10.0 (a), 12.0 (b), and 14.0 (c).}
\end{figure}

For strong magnetic fields, $h \gg \hbar v_F/\vert d_1 - d_2
\vert$, the integral over $dx$ in Eq.~(\ref{supercur3}), which
corresponds to summing over all possible trajectories in the
ferromagnets can be calculated in the saddle-point approximation.
As a result we find the current-phase relation
\begin{eqnarray} \label{supercur4}
j & = & 2 j_0\sqrt{\frac{v_F \hbar}{\pi h|d_1 - d_2|}}
\sum_{k=1}^{\infty} \frac{(-1)^{k+1}}{k^{3/2}} \nonumber \\
& \times & \cos \left( \frac{2kh|d_1 - d_2|}{v_F \hbar} +
\frac{\pi}{4} \right) \sin k\varphi.
\end{eqnarray}
We note, first of all, that the {\em amplitude} of the
supercurrent oscillations as a function of $\varphi$ decreases
algebraically with the exchange field, as $\sqrt{\hbar v_F/ h\vert
d_1 - d_2 \vert}$. This is a direct consequence of the fact that
we summed over all possible trajectories, and hence averaged over
the different phases acquired during propagation in the
ferromagnetic domains along these trajectories. Secondly, as far
as the {\em phase-dependence} of the supercurrent is concerned, it
is in general neither sinusoidal, nor saw-tooth-like. In
Fig.~\ref{fig2}, we plot $j(\varphi)$ for various values of
$h\vert d_1 - d_2 \vert /\hbar v_F \sim$ 10, such that the
saddle-point approximation is reasonable. We see that, as a
function of the exchange field, the supercurrent changes sign at a
given phase difference. Thus, depending on the parameter $h\vert
d_1 -d_2 \vert/\hbar v_F$, the junction either favors a 0-state or
a $\pi$-state. We finally note that for $d_2 = 0$,
Eqs.~(\ref{supercur3}) and (\ref{supercur4}) give the supercurrent
for a (single-domain) clean long SFS junction. This is, to our
knowledge, a new result as well. It implies in particular that a
clean SFS junction can also be a $\pi$-junction, in accordance
with previous results for different types of SFS hybrid
structures.

The important conclusion for the general case is that for a two-domain
contact the result is exactly the same as for a SFS junction with the
thickness $d_1 + d_2$ and the {\em effective} exchange field $h_{ef} =
h\vert d_1 - d_2\vert/(d_1 + d_2)$. In particular, if the thicknesses
are the same, $d_1 = d_2$, the magnetic field drops out -- we obtain
the sawtooth current-phase relation (\ref{supSNS1}) like for a SNS 
contact. In the language of Eilenberger equations, this statement is
obvious: Indeed, the only quantity sensitive to the magnetic field is
the phase $\alpha$ accumulated along the trajectory. Since each
trajectory is a straight line, each layer contributes with a weight
proportional to its thickness and with the sign depending on the
direction of the exchange field. This result is readily generalized to
the case of  many ferromagnetic layers in the antiparallel
configuration~\cite{foot2}.

\section{Disorder averaging}

Now we discuss how our two main observations for the supercurrent
--- power-law decay with magnetic field and the independence on
the magnetic field in the symmetric case $d_1 = d_2$ --- react to
the presence of disorder. Before performing this difficult task in
Section~\ref{section4} by solving the Usadel equations, we try to
use an easy way to understand the effect of impurities in this
Section. We introduce randomness in the thicknesses of the layers
(surface randomness). This simple and transparent calculation
provides us with results which are clear qualitative predictions
to be compared with the conclusions extracted from a more
complicated analysis of the Usadel equations.

We start from Eq.~(\ref{supercur3}), and imagine that the
interfaces as presented in Fig.~\ref{fig1} are not straight, but
exhibit small fluctuations in position. Since there is no
scattering at the interfaces, the only effect of such fluctuations
is that the thicknesses of the layers become random variables, and
the supercurrent~(\ref{supercur3}) must be averaged with respect to
this randomness. Let us take a Gaussian distribution for the
difference $d_1 - d_2$,
\begin{equation} \label{Gauss}
P(d_1 - d_2) = \frac{1}{\sqrt{\pi a}} \exp \left( - \frac{(d_1 -
d_2 - \bar d_1 + \bar d_2)^2}{a^2} \right),
\end{equation}
where $a \ll \bar d_1, \bar d_2$ has the meaning of a typical
scale of the interface fluctuations, and $\bar d_i$ are the
averaged values both thicknesses. Averaging Eq.~(\ref{supercur3}),
we obtain
\begin{eqnarray} \label{aver1}
\bar j & = & \frac{4 \bar{j}_0}{\pi}  \sum_{k=1}^{\infty}
\frac{(-1)^{k+1}}{k} \sin k\varphi \int_1^{\infty} \frac{dx}{x^3
\sqrt{x^2-1}} \nonumber \\
& \times & \exp \left( -\frac{k^2 h^2 a^2 x^2}{v_F^2\hbar^2}
\right) \cos \frac{2kh(\bar d_1- \bar d_2)x}{v_F\hbar},
\end{eqnarray}
where $\bar j_0 = \pi ev_F^2 \nu\hbar/(\bar d_1 + \bar d_2)$. In
strong fields, $h \gg \hbar v_F/a \gg \hbar v_F/(\bar d_1 - \bar
d_2)$, the integral is calculated in the saddle-point
approximation, and only the term with $k=1$  survives,
\begin{eqnarray} \label{aver2}
\bar j & = & - 2 \bar j_0 \sqrt{\frac{v_F \hbar}{\pi h \vert \bar
d_1 - \bar d_2\vert }}
 \exp \left( - \frac{h^2
a^2}{v_F^2\hbar^2} \right) \nonumber \\
& \times & \cos \left( \frac{2h(\bar d_1 - \bar d_2)}{v_F\hbar} +
\frac{\pi}{4} \right) \sin \varphi.
\end{eqnarray}
Thus, the averaging procedure brings out two, qualitatively new
features: (i) at high fields, the current-phase relation becomes
sinusoidal; (ii) the amplitude of the supercurrent oscillations
decays exponentially, rather than algebraically with $h$. In
addition, the exchange field still modulates the phase of the
oscillations, and can drive the contact to a $\pi$-state. The
property (i) stems from phase-averaging over diffusive
trajectories and is a common feature of all long disordered SNS
junctions ({\textit cf} Ref.~\onlinecite{Svidzinsky2}).

Eq.~(\ref{aver2}) does not apply to the symmetric case $\bar d_1 =
\bar d_2$. In this 
situation, for $h \gg v_F/a$, we have
\begin{equation}\label{aver3}
\bar j = -\frac{2 \bar j_0}{\sqrt{\pi}} \frac{v_F \hbar}{h a} \exp
\left( - \frac{h^2 
a^2}{v_F^2\hbar^2} \right) \sin \varphi.
\end{equation}
We see that even for the symmetric case the exponential dependence on
magnetic field persists. It reflects the fact that a quasiparticle
moving along a single trajectory spends in general unequal times in
both layers, and thus the contribution of each trajectory is magnetic
field dependent. However, there is no additional oscillating factor
due to the magnetic field: a symmetric junction is never in the
$\pi$-state. These features are confirmed qualitatively in the next
Section, where we analyze the behavior of a symmetric diffusive SFFS
junction, using the Usadel equations.

\section{Disordered SFFS contact from Usadel equations}
\label{section4}

We now consider a diffusive SFFS junction in the symmetric case
$d_1 = d_2 = d/2$. The junction is again assumed to be long, $d
\gg (\hbar D/\Delta)^{1/2}$, with $D$ being the diffusion
coefficient.

If the exchange magnetic energy does not exceed the inverse
elastic scattering time, $h \ll \hbar/\tau$, the Green's function
is almost isotropic, and the system can be described by Usadel
equations,
\begin{eqnarray} \label{usadel1}
& & D \partial_z \left[ F_{\sigma} \partial_z F^+_{\sigma} -
F_{\sigma}^+\partial_z F_{\sigma} \right] = 2
(\tilde\Delta F^+_{\sigma} - \tilde\Delta^* F_{\sigma}), \\
& & D \partial_z \left[ G_{\sigma} \partial_z F_{\sigma} -
F_{\sigma}
\partial_z G_{\sigma} \right] = 2 (\omega \pm ih\sigma) F_{\sigma} -
2\tilde\Delta G_{\sigma},  \nonumber
\end{eqnarray}
with the constraint $G_{\sigma}^2 + F_{\sigma}F^+_{\sigma} = 1$.
Here, as usual~\cite{Svidzinsky2},
\begin{displaymath}
\breve{G}_{\sigma} (\bm{r}) = \int d\bm{n} \ \breve{g}_{\sigma}
(\bm{r}, \bm{n}),
\end{displaymath}
and it actually only depends on the distance $z$ from the
ferromagnet-ferromagnet interface (Fig. \ref{fig1}). The
upper/lower signs describe the regions $-d/2 < z < 0$ and $0 < z <
d/2$, respectively, and $\tilde \Delta = i\Delta \exp(i\chi)$ in
the superconductors. In the following, we suppress the spin index
$\sigma$ where it does not lead to ambiguities.

Following Ref.~\onlinecite{Svidzinsky}, we solve the constraint
by introducing two complex-valued fields $\theta$ and $\eta$,
\begin{equation} \label{substit}
G = \cos \theta, \ \ \ F = \sin \theta e^{i \eta}, \ \ \ F^+ =
\sin \theta e^{-i\eta}.
\end{equation}
The equation for $\eta$ in the ferromagnets becomes
\begin{equation} \label{usadel2}
\partial_z (\eta' \sin^2 \theta) = 0,
\end{equation}
with the boundary conditions $\eta (\pm d/2) = \mp \varphi/2$. The
first integral yields
\begin{equation} \label{usadel3}
\eta' = \frac{I}{\sin^2 \theta}.
\end{equation}
where $I$ is an unknown constant. The current is expressed via
this constant,
\begin{eqnarray} \label{curdif1}
j & = & \frac{i\pi e D \nu}{2} \sum_{\sigma} T \sum_{\omega}
\left[ F_{\sigma} \partial_z F^+_{\sigma} - F^+_{\sigma}
\partial_z
F_{\sigma} \right] \nonumber \\
& = & \pi e D \nu \sum_{\sigma} T \sum_{\omega} I_{\sigma}.
\end{eqnarray}
To ensure the current conservation, $I$ must be the same in both
ferromagnetic layers. It is important, however, that we {\em do
not} assume that the current is conserving -- it follows naturally
from the consistency of our solution.

Using Eq.~(\ref{usadel2}), we also write the equation for $\theta$
in ferromagnets,
\begin{equation} \label{usadel4}
D\theta'' = D I^2 \frac{\cos \theta}{\sin^3 \theta} + 2 (\omega
\pm i h \sigma) \sin \theta,
\end{equation}
with the first integral
\begin{equation} \label{usadel5}
D\theta'^2 = -\frac{D I^2}{\sin^2 \theta} - 4(\omega \pm ih\sigma)
\sin \theta + \mbox{const}.
\end{equation}
Now, for the long junctions, the boundary conditions for $\theta$
at $z = \pm d/2$ are essentially the same they would be at the
interface between a semi-infinite superconductor and a
semi-infinite ferromagnet. To find these boundary conditions, we
write the corresponding equation for the superconductors,
\begin{displaymath}
D\theta'^2 = - 4 \omega \sin \theta - 4\Delta \cos \theta +
\mbox{const},
\end{displaymath}
Taking into account that in the bulk superconductor $\theta =
\pi/2$, in the bulk ferromagnet $\theta = 0$, and requiring the
continuity of $\theta$ and $\theta'$ at the interface, we obtain
the following boundary conditions,
\begin{displaymath} \label{bound2}
\omega \pm ih\sigma (1 - \cos \theta) + \Delta (1 - \sin \theta) +
\frac{I^2}{\sin^2 \theta} = 0,
\end{displaymath}
at $z = \mp d/2$. Although our equations describe the behavior of an
SFFS junction for an arbitrary relation between $h$ and $\Delta$, we
concentrate in the following on the case $T, h \ll \Delta$. As we show
below, in this situation the current $I$ is exponentially small, and
the boundary condition for $\theta$ reduce to $\theta(z = \pm d/2) =
\pi/2$.

Since the Usadel equations posses obvious symmetries
$\theta_{\sigma} (\omega) = \theta_{-\sigma} (\omega) + \pi$,
$\eta_{\sigma} (\omega) = \eta_{-\sigma} (\omega) + \pi$, in the
sequel we only consider $\omega
> 0$.

The field $\theta$ must rapidly decay away from superconductors
and stay exponentially small within the ferromagnets. We start
first solving Eq.~(\ref{usadel5}) at $z \ll d$, where $\theta \ll
1$, and the trigonometric functions can be expanded. Then
Eq.~(\ref{usadel5}) can be integrated. The solution is too
cumbersome to be written down here, its asymptotics for $\vert z
\vert \to \infty$ are
\begin{eqnarray} \label{usadel6}
\theta & = & \frac{1}{2} \sqrt{\left[ \theta_0 +
\sqrt{\frac{D\gamma^2}{2(\omega \pm ih\sigma)}}\right]^2 +
\frac{DI^2}{2\theta_0^2 (\omega \pm ih\sigma)}} \nonumber \\
& \times & \exp \left( \sqrt{\frac{2}{D}} (\alpha + i\beta \sigma)
\vert z \vert \right),
\end{eqnarray}
with the notations $\theta_0 = \theta (z = 0)$, $\gamma = \theta'
(z = 0)$, and
\begin{displaymath}
\alpha = \frac{1}{\sqrt{2}} \sqrt{\sqrt{\omega^2 + h^2} + \omega};
\ \ \ \beta = \pm \frac{1}{\sqrt{2}} \sqrt{\sqrt{\omega^2 + h^2} -
\omega}.
\end{displaymath}

Next, we solve Eq.~(\ref{usadel5}) close to the interfaces, $\vert
z - d/2 \vert \ll d$. We assume that $I/\theta_0$, $\gamma$ are
both exponentially small (to be checked later) and obtain
\begin{equation} \label{usadel7}
\tan \frac{\theta}{4} = \tan \frac{\pi}{8} \exp \left(
\sqrt{\frac{2}{D}} (\alpha + i\beta \sigma) (\vert z \vert -
\frac{d}{2}) \right).
\end{equation}
Far from the interface, $\theta \ll 1$, the solution becomes
exponential. Matching the exponential asymptotics of
Eqs.~(\ref{usadel6}) and (\ref{usadel7}), we find the condition
\begin{eqnarray} \label{usadel8}
& & \left[ \theta_0^2 \mp \sqrt{\frac{D\gamma^2}{2(\omega \pm
ih\sigma)}} \right]^2 + \frac{DI^2}{2\theta_0^2 (\omega \pm
ih\sigma)}
\nonumber \\
& & = 64 \tan^2 \frac{\pi}{8} \exp \left( - \sqrt{\frac{2}{D}}
(\alpha + i\beta \sigma) d \right).
\end{eqnarray}

We now integrate Eq.~(\ref{usadel3}). Since $\theta(x)$ grows
exponentially away from $x=0$, the sine in the denominator can be
replaced by its argument. We then find 
\begin{equation} \label{usadel9}
\sqrt{\frac{DI^2}{2(\omega \pm ih\sigma)}} = \theta_0 \left(
\theta_0 \mp \sqrt{\frac{D\gamma^2}{2(\omega + ih\sigma)}} \right)
\tan \left( \frac{\varphi}{2} \pm \eta_0 \right),
\end{equation}
with $\eta_0 = \eta(0)$. We proceed by calculating the four quantities
$I$, $\theta_0$, $\eta_0$, and $\gamma$. The result is
\begin{eqnarray} \label{curus1}
I & = & \frac{64 (\sqrt{2} - 1)^2}{\sqrt{D}} \sqrt{\frac{\omega^2
+
h^2}{\sqrt{\omega^2 + h^2} + \omega}} \nonumber \\
& \times & \exp \left( - \sqrt{\frac{2}{D}} \alpha d \right) \sin
\varphi.
\end{eqnarray}
Note that $I$ does not depend on $\sigma$. It can be easily
checked that $I/\theta_0$ and $\gamma$ are exponentially small,
which justifies the approximations we have made to arrive at
Eq.~(\ref{curus1}).

Now we calculate the supercurrent according to
Eq.~(\ref{curdif1}). For high temperatures $T \gg \hbar D/d^2, h$
only the term with $\omega = \pi T$ is important, and we obtain
\begin{eqnarray} \label{surdif2}
j & = & \sqrt{2} j_{0,{\rm diff}} \left( \frac{\pi k_BT d^2}{\hbar
D }\right)^{3/2}
\nonumber \\
& \times & \exp \left( -\frac{d}{\sqrt{\hbar D}} \sqrt{\pi T +
\sqrt{h^2 + \pi^2 T^2}} \right) \sin \varphi,
\end{eqnarray}
where we introduced $j_{0,{\rm diff}} = 128 (\sqrt{2} - 1)^2 e\nu
\hbar D^2/d^3$. In high magnetic fields $h \gg T, \hbar D/d^2$ the
terms with $\omega < h$ contribute,
\begin{equation} \label{surdif3}
j = j_{0,{\rm diff}} \left(\frac{h d^2}{\hbar D}\right)^{3/2} \exp
\left( - \sqrt{\frac{h}{\hbar D}} d \right) \sin \varphi.
\end{equation}
We note the two main features of the solution in the diffusive
case. First, the current-phase relation is sinusoidal. This
corresponds to the result for the long diffusive SNS
contact~\cite{Svidzinsky2}. Then, the supercurrent decays
exponentially with magnetic field, in contrast to the power-law
decay in the clean case.

Similarly, we can treat a single-layer SFS junction of a thickness
$d$. The result for $h \gg \hbar D/d^2, T$  reads
\begin{eqnarray} \label{surdif4}
j  & = &  j_{0,{\rm diff}} \left(\frac{h d^2}{\hbar
D}\right)^{3/2} \nonumber \\
& \times & \exp \left( - \sqrt{\frac{h}{\hbar D}} d \right)
 \sin \left( \sqrt{\frac{h}{\hbar D}} d \right) \sin
\varphi.
\end{eqnarray}
Thus, comparing Eq.~(\ref{surdif4}) with Eq.~(\ref{surdif3}) we see
that a long diffusive SFS contact can be a $\pi$-junction, depending
on the thickness of the ferromagnet, whereas a similar symmetric SFFS
contact with anti-parallel configuration of the domains is not a
$\pi$-junction.

\section{Discussion}

We considered the behavior of the supercurrent in long SFS
junctions. We obtained new expressions for single-domain ballistic and
diffusive contacts and confirmed that the $0$ to $\pi$ transition can
be induced in these systems. However, our main focus is on 
the situation when the ferromagnetic region is split into two
ferromagnetic domains with equal but opposite magnetization. In the
ballistic case, this system behaves as a single-domain SFS junction,
with the effective exchange field $h_{ef} = h \vert d_1 - 
d_2\vert/(d_1 + d_2)$. Such a system exhibits a non-sinusoidal
current-phase relation, and a power-law decay of the supercurrent with
thickness and exchange field. If the thicknesses of the both domains
are the same, the effective field vanishes. Disorder, considered both
as geometrical fluctuations of the thickness, or randomly positioned 
impurities, restores exponential decay and sinusoidal phase dependence
of the supercurrent. A system with two domains of the same width is
never in the $\pi$-state.

To obtain these results, we made a number of simplifying
assumptions. The superconductor-ferromagnet interfaces, as well as the 
boundary between the two ferromagnetic domains, are assumed to be
ideal (no scattering) and sharp. This can be realized in multilayered
structures, where the ferromagnetic layers can be artificially
constructed and kept very clean. Another, more attractive option, is
real ferromagnetic domains. A domain wall has a finite width,
typically of order of the mean free path, or wider. This induces
reflection of electrons from the domain wall, and additionally
generates the triplet pairing between electrons and holes. These 
factors need to be taken into account for a quantitative comparison
between theory and experiment. However, we do not expect them to add
qualitatively new features into the picture we presented.

\section*{ Acknowledgments} We thank J.~Aarts, E.~V.~Bezuglyi,
Yu.~M.~Galperin, A.~A.~Golubov, W.~Guichard, and Yu.~V.~Nazarov for
useful discussions. FH acknowledges the hospitality of Delft
University of Technology. This work was supported by the Netherlands
Foundation for Fundamental Research on Matter (FOM) and by Institut
Universitaire de France.


\begin{thebibliography}{100}


\bibitem{Bulaevskii77} L.~N.~Bulaevskii, V.~V.~Kuzii, and
A.~A.~Sobyanin, Pis'ma Zh. \'Eksp. Teor. Fiz. {\bf 25}, 314 (1977)
[JETP Lett. {\bf 25}, 290 (1977)].


\bibitem{Bulaevskii} A.~I.~Buzdin, L.~N.~Bulaevskii, and
S.~V.~Panyukov, Pis'ma Zh. \'Eksp. Teor. Fiz. {\bf 35}, 147 (1982)
[JETP Lett. {\bf 35}, 178 (1982)].


\bibitem{Radovic00} L.~Dobrosavljevi\'c-Gruji\'c, R.~Ziki\'c, and
Z.~Radovi\'c, Physica C {\bf 331}, 254 (2000).


\bibitem{Chtchelkatchev} N.~M.~Chtchelkatchev, W.~Belzig,
Yu.~V.~Nazarov, and C.~Bruder, Pis'ma Zh. \'Eksp. Teor. Fiz. {\bf
74}, 357 (2001) [JETP Lett. {\bf 74}, 323 (2001)].


\bibitem{Bergeret2} F.~S.~Bergeret, A.~F.~Volkov, and K.~B.~Efetov,
Phys. Rev. B {\bf 64}, 134506 (2001).


\bibitem{Buzdin} A.~I.~Buzdin and M.~Yu.~Kupriyanov, Pis'ma
Zh. \'Eksp. Teor. Fiz. {\bf 53}, 308 (1991) [JETP Lett. {\bf 53},
321 (1991)]; A.~I.~Buzdin, B.~Vuji\v{c}i\'c, and
M.~Yu.~Kupriyanov, Zh. \'Eksp. Teor. Fiz. {\bf 101}, 231 (1992)
[Sov. Phys. JETP {\bf 74}, 124 (1992)].


\bibitem{Demler} E.~A.~Demler, G.~B.~Arnold, and M.~R.~Beasley,
Phys. Rev. B {\bf 55}, 15174 (1997).


\bibitem{Heikkila} T.~T.~Heikkil\"a, F.~K.~Wilhelm, and G.~Sch\"on,
Europhys. Lett. {\bf 51}, 434 (2000).


\bibitem{Tanaka} Y.~Tanaka and S.~Kashiwaya, Physica C {\bf 274}, 357
(1997).


\bibitem{Fogelstrom} M.~Fogelstr\"om, Phys. Rev. B {\bf 62}, 11812
(2000); J.~C.~Cuevas and M.~Fogelstr\"om, Phys. Rev. B {\bf 64},
104502 (2001).


\bibitem{Bergeret1} F.~S.~Bergeret, A.~F.~Volkov, and K.~B.~Efetov,
Phys. Rev. Lett. {\bf 86}, 3140 (2001).


\bibitem{Barash1} Yu.~S.~Barash and I.~V.~Bobkova, Phys. Rev. B {\bf
65}, 144502 (2002).


\bibitem{Barash2} Yu.~S.~Barash, I.~V.~Bobkova, and T.~Kopp,
Phys. Rev. B {\bf 66}, 140503 (2002).


\bibitem{Koshina1} E.~A.~Koshina and V.~N.~Krivoruchko, Pis'ma
Zh. \'Eksp. Teor. Fiz. {\bf 71}, 182 (2000) [JETP Lett. {\bf 71},
123 (2000)]; E.~Koshina and V.~Krivoruchko, Phys. Rev. B {\bf 63},
224515 (2001).


\bibitem{Koshina2} V.~N.~Krivoruchko and E.~A.~Koshina, Phys. Rev. B
{\bf 64}, 172511 (2001).


\bibitem{Krivoruchko} V~N.~Krivoruchko and R.~V.~Petryuk, Phys. Rev. B
{\bf 66}, 134520 (2002).


\bibitem{Golubov1} A.~A.~Golubov, M.~Yu.~Kupriyanov, and
Ya.~V.~Fominov, Pis'ma Zh. \'Eksp. Teor. Fiz. {\bf 75}, 223 (2002)
[JETP Lett. {\bf 75}, 190 (2002)].


\bibitem{Golubov2} A.~A.~Golubov, M.~Yu.~Kupriyanov, and
Ya.~V.~Fominov, Pis'ma Zh. \'Eksp. Teor. Fiz. {\bf 75}, 709 (2002)
[JETP Lett. {\bf 75}, 588 (2002)].


\bibitem{Radovic03} Z.~Radovi\'c, N.~Lazarides, and N.~Flytzanis,
cond-mat/0305437.


\bibitem{Ryazanov} V.~V.~Ryazanov, V.~A.~Oboznov, A.~Yu.~Rusanov,
A.~V.~Veretennikov, A.~A.~Golubov, and J.~Aarts, Phys. Rev. Lett.
{\bf 86}, 2427 (2001).


\bibitem{Kontos} T.~Kontos, M.~Aprili, J.~Lesueur, F.~Gen\^et,
B.~Stephanidis, and R.~Boursier, Phys. Rev. Lett. {\bf 89}, 137007
(2002).


\bibitem{Guichard} W.~Guichard, M.~Aprili, O.~Bourgeois, T.~Kontos,
J.~Lesueur, and P.~Gandit, Phys. Rev. Lett. {\bf 90}, 167001
(2003).


\bibitem{Blum} Y.~Blum, A.~Tsukernik, M.~Karpovski, and A.~Palevski,
cond-mat/0203408.


\bibitem{Radovic99} Z.~Radovi\'c, L.~Dobrosavljevi\'c-Gruji\'c, and
B. Vuji\v{c}i\'c, Phys. Rev. B {\bf 60}, 6844 (1999).


\bibitem{Larkin} F.~S.~Bergeret, K.~B.~Efetov, and A.~I.~Larkin,
Phys. Rev. B {\bf 62}, 11872 (2000).


\bibitem{Bergeret3} F.~S.~Bergeret, A.~F.~Volkov, and K.~B.~Efetov,
Phys. Rev. Lett. {\bf 86}, 4096 (2001); Phys. Rev. B {\bf 65},
134505 (2002).


\bibitem{Kadigrobov} A.~Kadigrobov, R.~I.~Shekhter, and M.~Jonson,
Europhys. Lett. {\bf 54}, 394 (2001).


\bibitem{Spivak} A.~Yu.~Zyuzin, B.~Spivak, and M.~Hru\v{s}ka,
cond-mat/0204123.


\bibitem{Melin} R.~Melin, J. Phys. Cond. Matter {\bf 13}, 6445
(2001); N.~M.~Chtchelkatchev and I.~S.~Burmistrov,
cond-mat/0303014.


\bibitem{foot1} Generalization to three dimensions (planar contact) is
straightforward and the results are qualitatively the same.


\bibitem{Svidzinsky} A.~V.~Svidzinskii, {\em Spatially Non-homogeneous
Problems of Superconductivity}, Nauka, Moscow (1982).


\bibitem{Ishii} C.~Ishii, Prog. Theor. Phys. {\bf 44}, 1525 (1970);
J.~Bardeen and J.~L.~Johnson, Phys. Rev. B {\bf 5}, 72 (1972);
A.~I.~Bezuglyi, I.~O.~Kulik, and Yu.~N.~Mitsai, Fiz. Nizkikh Temp.
{\bf 1}, 57 (1975) [Sov. J. Low Temp. Phys. {\bf 1}, 27 (1975)].


\bibitem{foot2} In this case, the magnetic fields in all the layers
need to be collinear, otherwise triplet pairing is generated.


\bibitem{Svidzinsky2} A.~I.~Makeev and A.~V.~Svidzinskii,
Teor. Mat. Fiz. {\bf 44}, 85 (1980) [Theor. Math. Phys. {\bf 44},
617 (1980)]; A.~D.~Zaikin and G.~F.~Zharkov, Fiz. Nizkikh Temp. {\bf
7}, 375 (1981) [Sov. J. Low Temp. Phys. {\bf 7}, 184 (1981)].


\end{thebibliography}
\end{document}